\documentclass[a4paper]{article}

\usepackage{INTERSPEECH2022}
\usepackage{multirow,makecell}
 \usepackage{xcolor}
\usepackage[ruled,linesnumbered]{algorithm2e}
\title{Iterative Sound Source Localization for Unknown Number of Sources}
\name{Yanjie Fu$^{1}$, Meng Ge$^{1,2,*}$, Haoran Yin$^{1}$, Xinyuan Qian$^{2}$, Longbiao Wang$^{1,*}$, \\ Gaoyan Zhang$^{1}$, Jianwu Dang$^{1,3}$}
\address{
  $^{1}$Tianjin Key Laboratory of Cognitive Computing and Application,\\
  College of Intelligence and Computing, Tianjin University, Tianjin, China\\
  $^2$Department of Electrical and Computer Engineering, National University of Singapore, Singapore \\
  $^{3}$Japan Advanced Institute of Science and Technology, Ishikawa, Japan}
\email{$\{$fuyanjie, gemeng, haoran\_yin, longbiao\_wang$\}$@tju.edu.cn}

\begin{document}

\maketitle
\begin{abstract}
Sound source localization aims to seek the direction of arrival (DOA) of all sound sources from the observed multi-channel audio. For the practical problem of unknown number of sources, existing localization algorithms attempt to predict a likelihood-based coding (i.e., spatial spectrum) and employ a pre-determined threshold to detect the source number and corresponding DOA value. However, these threshold-based algorithms are not stable since they are limited by the careful choice of threshold. To address this problem, we propose an iterative sound source localization approach called ISSL, which can iteratively extract each source's DOA without threshold until the termination criterion is met. Unlike threshold-based algorithms, ISSL designs an active source detector network based on binary classifier to accept residual spatial spectrum and decide whether to stop the iteration. By doing so, our ISSL can deal with an arbitrary number of sources, even more than the number of sources seen during the training stage. The experimental results show that our ISSL achieves significant performance improvements in both DOA estimation and source number detection compared with the existing threshold-based algorithms.

\end{abstract}

\section{Introduction}

Sound source localization (SSL) is the technique of estimating the position of one or several sound sources from the multi-channel signals captured by the microphone array. It can be simplified to the estimation of direction-of-arrival (DOA) to the microphone array of each sound source. This localization cue is now an imperative part of many real-world speech applications such as speech separation \cite{chazan2019multi, DBLP:journals/corr/abs-2107-06853,li2021MIMO}, speaker extraction \cite{ge2022spex,zhang2021ADL,xu2021GRNN}, speech recognition \cite{kim2015recognition,subramanian2022recognition} and speaker diarization \cite{ajmera2004clustering,otterson2007improved,zheng2021real}, which makes SSL a valuable research topic.

In the past few decades, most researchers have focused on DOA estimation with a known number of sound sources. They usually borrow the nonlinear ability of neural network to derive source's direction from a low-level or high-level signal representation, including waveform \cite{suvorov2018deep,vecchiotti2019end,bologni2021acoustic}, spectrogram \cite{adavanne2018direction,he2021neural}, acoustic intensity vector \cite{liu2021deep,perotin2019crnn,Perotin2018CRNNbasedJA} and GCC-PHAT \cite{xiao2015learning,he2018deep,li2018online}. These prior algorithms have laid the foundation for follow-up SSL studies. Despite their promising results, localization of sound sources in real-world scenarios with unknown number of sources is still a challenge.


Recently, a lot of works utilize a predefined threshold to deal with the DOA estimation problem when the number of active sources is not known in advance, including \cite{chazan2019multi,adavanne2018direction,he2018deep,7846325}. For example, He et al. \cite{he2018deep} design a 4-layer fully connected network and a 4-layer convolution neural network to map the GCC-PHAT input feature to desired spatial spectrum coding. In addition, He et al. \cite{he2019adaptation} further the idea and derive the spatial spectrum coding from both the real and imaginary parts of the complex-valued spectrogram.
These threshold-based solutions aim to predict this likelihood-based spatial spectrum output at each possible azimuth direction, and then employ an empirically-defined threshold to detect the source number and estimate the corresponding DOA value. However, these methods are extremely dependent on the threshold selection and are not stable in real-world applications, since real-world scenes often contain various noises and different sound source types.

\begin{figure*}[t] 
\centering 
\includegraphics[width=17cm]{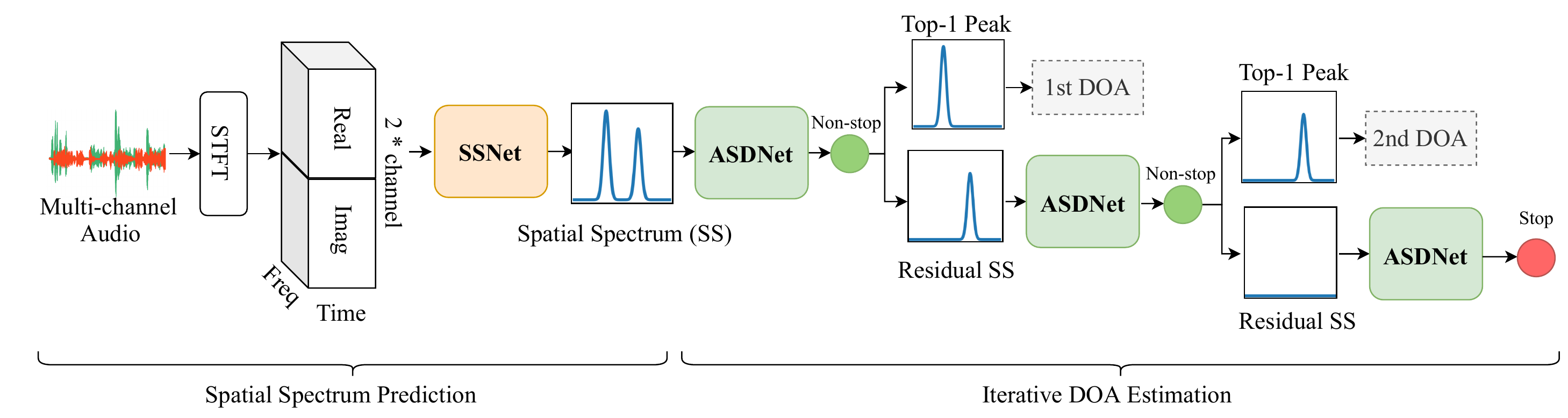} 
\caption{Illustration of our proposed sound source localization framework ISSL when the number of sources is 2.  SSNet processes the multi-channel audio feature and produces one spatial spectrum. ASDNet predicts if there still exists active sound source. Iterative DOA Estimation predicts the final DOAs as well as the number of active sound sources. 
}
\label{Fig.Overview} 
\end{figure*}

To address the threshold problem for unknown number of sources, we propose an iterative sound source localization approach called ISSL (Fig. \ref{Fig.Overview}), which avoids the threshold assumption by using an iterative strategy and doesn't require the knowledge of the number of sources in the observed signal. Inspired by the success that spatial spectrum can encode a signal with an arbitrary number of sources into a likelihood-based vector, we first build a spatial spectrum prediction module using ResNet-based \cite{he2016deep} network (Fig. \ref{Fig.SSNet}(a)) to obtain the spatial spectrum output coding for the following DOA estimation. Unlike the existing threshold-based algorithms, we then design an iterative DOA estimation module using VGG-based \cite{simonyan2014very} network (Fig. \ref{Fig.SSNet}(b)) to accept spatial spectrum to iteratively extract each source's DOA until the termination criterion is met. This iterative DOA estimation module replaces the original threshold operation with a termination strategy designed by a binary classifier. By doing so, our ISSL system can handle an arbitrary number of sources without a pre-determined threshold, even more than the number of sources seen during training. Compared with the threshold-based algorithm on source number detection, our ISSL system naturally avoids the instability of the threshold, and the experimental results also verify that our system achieves 19.41\% and 16.95\% absolute improvement in detection accuracy for scenarios containing at most 3 and 4 sources, respectively.

\section{ISSL System}
In this section, we describe our proposed ISSL system for unknown multiple sound sources localization. Then we introduce how we infer with this system, followed by training methods.

\subsection{Overview}

Our ISSL system consists of spatial spectrum prediction module and iterative DOA estimation module, as shown in Fig. \ref{Fig.Overview}. The former module aims to predict the spatial spectrum from the multi-channel input audio, while the latter one counts the source number and estimates each source's DOA via the proposed iterative strategy. The whole training and inference process is data-driven and has no threshold operation.


\subsection{Spatial Spectrum Prediction}

\begin{figure*}[htb] 
\centering 
\includegraphics[width=16cm]{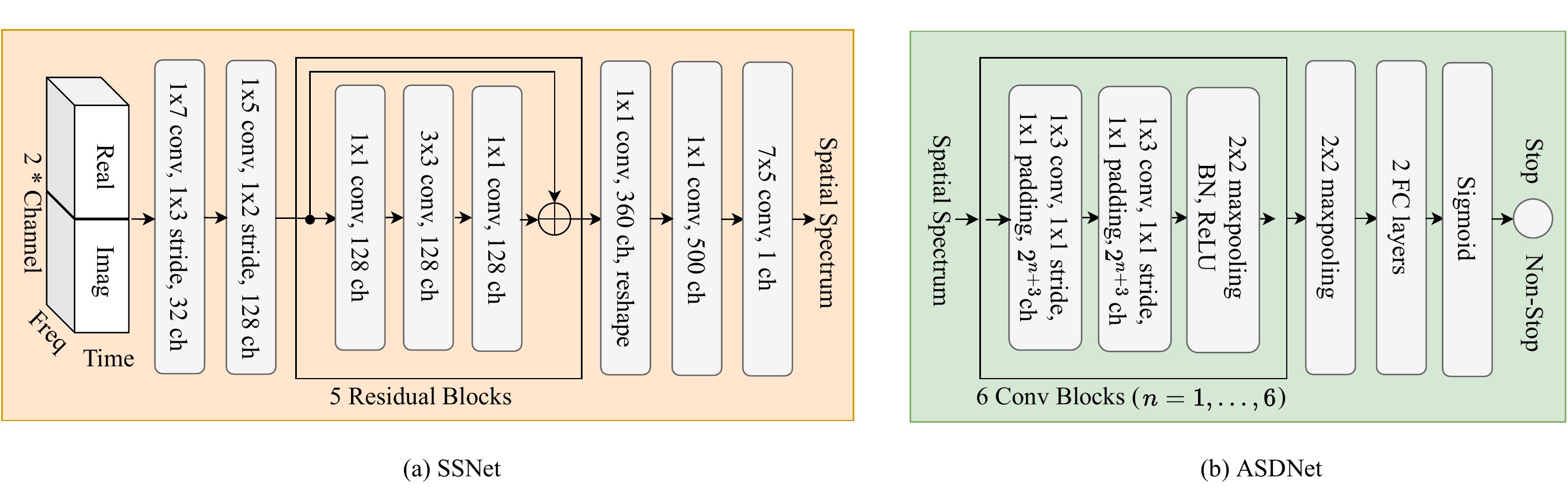} 
\caption{Architectures of Spatial Spectrum Network (SSNet) and Active Source Detector Network (ASDNet).} 
\label{Fig.SSNet} 
\end{figure*}

The spatial spectrum prediction module is built by using Spatial Spectrum Network (SSNet) (Fig. \ref{Fig.SSNet}(a)), which learns the non-linear mapping from the audio segment to the spatial spectrum.



Formally, let $X_c$ be the complex STFT of the $c$-th channel mixture signal $x_c$ of one segment.
We concatenate the real part and imaginary part of STFT along the channel axis to obtain the feature representation $v$ which is defined as:
\begin{equation}
v = \{\mathcal{R}e[X],\mathcal{I}m[X]\}
\end{equation}
where $X=\{X_c\}_{c=1}^C$ is the input complex-value spectrogram computed from all $C$-microphones. After that, we use $v$ as the input of SSNet to predict the spatial spectrum:
\begin{equation}
\hat{p}=\text{SSNet}(v)
\end{equation}
Specifically, the network structure of SSNet is shown in Fig. \ref{Fig.SSNet}(a). It contains two convolutional layers for downsampling in the frequency dimension, five residual blocks for the high-level feature extraction, one $1\times1$ convolutional layer for projection onto the DOA space followed by an operation of swapping the DOA and the frequency axes. The last two layers convolve along the time and DOA axes and the final output representation is a 360-dimensional vector.

Inspired by \cite{he2018deep}, we adopt the likelihood-based spatial spectrum to represent the probability of sound source locating at the 360 individual directions. Specifically, each element of the encoded 360-dimensional vector $p(\theta_i)$ is assigned to a particular azimuth direction $\theta_i \in \{1^\circ,2^\circ,...,360^\circ\}$. The values of the vector follow a Gaussian distribution that maximizes at the ground truth DOA, which is defined as follow:
\begin{equation}
p(\theta_i)=\begin{cases}\max _{\theta^{\prime} \in y}\left\{\exp \left(-\frac{d\left(\theta_i,\theta^{\prime}\right)^{2}}{\sigma^{2}}\right)\right\} & \text { if }|y|>0 \\ 0 & \text { otherwise }\end{cases}
 \label{encoding}
\end{equation}
where $y$ is the set of the ground truth sound sources, $|y|$ is the number of sources, $\theta^{\prime}$ is the ground truth DOA of one source, $\sigma$ is a predefined constant that controls the width of the Gaussian function and $d(\cdot,\cdot)$ denotes the angular distance. 

We train SSNet by computing the mean squared error (MSE) between the segment-wise output spatial spectrum $\hat{p}$ from SSNet and the ground truth $p$, which represented as
\begin{equation}
\mathcal{L}_{\text{SSNet}}=||\hat{p}-p||^2
\end{equation}

\subsection{Iterative DOA Estimation}
In practical sound source localization scenarios, we will not be informed of the source number beforehand. Thus, we have no idea how many peaks in the spatial spectrum should be taken as predictions during inference. In \cite{he2019adaptation}, the peaks in the predicted spatial spectrum above a given threshold are taken as predictions when the number of sources is unknown. This threshold-based method does achieve relatively good results of localization for unknown number of sources. However, the optimal threshold is hard to choose and a pre-determined threshold for all domains is prone to errors due to the domain shift problem.

To perform better localization for unknown number of sources, we propose to estimate DOAs by applying the Active Source Detector Network (ASDNet) iteratively inspired by \cite{takahashi2019recursive}. Instead of predicting all DOAs at once, the proposed approach predicts only one DOA from a spatial spectrum mixture at a time and the residual spatial spectrum is fed back to the model for the next iteration, as shown in Fig. 1. 

Specifically, we use ASDNet in the task of binary classification to estimate if there exists active sound source given a spatial spectrum. If so, we zero the highest peak of the spatial spectrum to separate the sound source from the spatial spectrum mixture and feed the new spatial spectrum into the model to continue the iteration. Otherwise, we stop the iteration and get a DOA estimation as well as an estimation of the source number. The complete algorithm is shown in Algorithm 1. 

\begin{algorithm}
\caption{Iterative DOA Estimation for Unknown Number of Sources}\label{algorithm1}
\SetKwInOut{Input}{Input}\SetKwInOut{Output}{Output}
\Input{One 360-dimensional predicted spatial spectrum $\hat{p}$}
\Output{Estimated DOAs $results$ \& number of sources $nos$}
$results\leftarrow \{\}$,
$nos \leftarrow 0 $

Find the maximum value of the spectrum and its corresponding DOA:
$\hat{\theta} = argmax_{\forall\theta}\ \hat{p}(\theta)$

Feed $\hat{p}$ into ASDNet

\While{ASDNet detects an active source}{


Update $\hat{p}$ by setting the values at $\hat{\theta}$ and its neighborhood to 0

$results\leftarrow results \cup \{\hat{\theta}\}$

$nos \leftarrow nos + 1 $

Find the maximum value of the spectrum and its corresponding DOA:
$\hat{\theta} = argmax_{\forall\theta}\ \hat{p}(\theta)$

Feed $\hat{p}$ into ASDNet
}
\Return{results, nos}
\end{algorithm}

In our ASDNet architecture presented in Fig. \ref{Fig.SSNet}(b), we insert a height dimension and a channel dimension into the input 360-dimensional spatial spectrum's shape, respectively. As a result, the input shape is (1, 1, 360). Our proposed ASDNet adopts VGG-like architecture \cite{simonyan2014very}, which consists of 12 convolutional layers and 2 fully connected (FC) layers. The 6 convolutional blocks have 16, 32, 64, 128, 256 and 512 filters, respectively. Each of the convolutional layers is followed by a Batch Normalization layer \cite{ioffe2015batch} and Rectified Linear Unit \cite{glorot2011deep}. The first hidden layer has 256 neurons. The second hidden layer is a FC layer with 32 neurons. The last output layer has only one neuron with sigmoid activation function. We label the residual spatial spectrum containing active source as 0 and 1 the other way around. During each iteration of inference, we interpret the output value above 0.5 as ``stop" and below 0.5 as ``non-stop". We train ASDNet to minimize the binary cross entropy between the label value and the output value as follows:
\begin{equation}
\mathcal{L}_{\text{ASDNet}}=-\frac{1}{N}\sum_{n=1}^{N}(z_{n}log\hat{z_n}+(1-z_{n})log(1-\hat{z_n}))
\end{equation}
where $N$ is the total amount of samples and $n=1,...,N$ is the sample index, $z_n\in\{0,1\}$ is the label value of the $n^{th}$ sample, $\hat{z_n}\in\{0,1\}$ is the model's output value of the $n^{th}$ sample.
\subsection{Network Training}
For training ISSL system, we need to train SSNet and ASDNet separately. Once finish training SSNet, we build a spatial spectrum dataset predicted by SSNet for further training ASDNet.

Specifically, we train SSNet with a simulated multi-channel audio dataset in the first step. Then, we train ASDNet with the spatial spectrum dataset containing samples inferred from a dataset which is the same duration as the training set using SSNet. Among the samples, the positive samples which indicate to stop the iteration are the residual spatial spectrum of the 1\textsuperscript{st} iteration's result of 1 source, 2\textsuperscript{nd} iteration's result of 2 sources and 3\textsuperscript{rd} iteration's result of 3 sources since no active sources remain in the residual spatial spectrum. All other combinations such as the 1\textsuperscript{st} iteration of 2 sources mixture are used as negative samples as the residual spatial spectrum still contain sound.

\begin{table}[t]
\centering
\small
\caption{The parameters settings of simulated rooms.}
    \setlength{\tabcolsep}{1.2mm}{
        \begin{tabular}{c|c|c|c}
        \toprule
        \textbf{Length}$(L)$ & \textbf{Width}$(W)$ & \textbf{Height}$(H)$ & \textbf{RT60} \\
        \midrule
        \lbrack 4m, 6m)  & \multirow{3}{*}{\lbrack 3m, $L$\rbrack } &  \multirow{3}{*}{\lbrack 2.7m, 3.5m\rbrack } & \lbrack 0.2, 0.5\rbrack  \\
        \lbrack 6m, 10m)  & & & \lbrack 0.3, 0.6\rbrack  \\
        \lbrack 10m, 15m\rbrack & & & \lbrack 0.4, 0.7\rbrack  \\
        \bottomrule
        \end{tabular}}
\vspace{-5pt}
\end{table}

\section{Experiment and Discussion}

\subsection{Dataset}

We simulate multi-channel mixture audio of different number of sources from 0 to 4. The clean speech audio data are randomly selected from the VCTK \cite{yamagishi2019cstr} corpus at a sample rate of 48kHz. The simulated database contains 109 speakers and is divided into three sets: training set (89 speakers), validation set (10 speakers), and test set (10 speakers). All multi-channel audio data are generated by convolving clean speech audio with simulated room impulse responses using Pyroomacoustics  \cite{scheibler2018pyroomacoustics}. 

Specifically, we first generate spatialized 4-channel audio data of single-source speech in cubic rooms of random sizes. We randomly place the microphone array and sound sources in the room. The distances between the sound source and the walls are at least 0.5m. The microphone array geometry is set according to that of the microphone array which is used during collecting SSLR database \cite{he2018deep}. Then, we respectively mix 3 and 4 randomly selected audios with a minimum angular spacing of 20 degree. We duplicate or truncate the original audio to the guarantee all mixed audios are 5 seconds long. We slice the audios into 170 ms long segments (8192 samples in 48kHz recordings) and label each segment with the active sources' DOAs. It is worth mentioning that we identify active sources in each segment with the assistance of SpeechBrain \cite{speechbrain} VAD results on every single source. Thus, we obtain VCTK-3mix and VCTK-4mix that contain 0,1,2,3 and 0,1,2,3,4 sources segments, respectively. The total duration of training set, validation set and test set are 27.8h, 2.78h and 2.78h, respectively. We simulated 50 rooms for training set and 2 rooms each for validation set and test set. The parameters of the simulated rooms are listed in Table 1. The dataset in our experiments can be found here\footnote{https://github.com/FYJNEVERFOLLOWS/VCTK-SIM}.
\subsection{Experimental Setup}
We train all models in our ISSL system for 30 epochs. Adam \cite{kingma2014adam} optimizer is used in training both SSNet and ASDNet. We apply early stopping to prevent the network from overfitting. We set the minibatch size to 100 in all experiments. The parameter $\sigma$ of Gaussian curve in spatial spectrum coding is 8. In our experiments, the segments are 170ms long 4-channel audio (8192 samples with 48kHz sampling rate). We compute the STFT of these segments with a 43ms (2048 samples) long window and 50\% long hop. Thus, we have 7 frames of STFT from each segment as the network input. We only use the frequency bins between 100 and 8000 Hz, so the number of frequency bins is reduced to 337. The dimension of the feature representation $v$ is $7 \times 337 \times 8$. Source code in this paper is available\footnote{https://github.com/FYJNEVERFOLLOWS/ISSL}.

\subsection{Comparative study on VCTK-3mix and VCTK-4mix}
We compare our ISSL system with the previous state-of-the-art threshold-based baseline systems on VCTK-3mix and VCTK-4mix in terms of Precision, Recall and F1-score \cite{1992F1Score}. From Table \ref{tbl:small_test_set}, we find that our ISSL system significantly outperforms the ``Threshold (CNN)'' and ``Threshold (Res)'' systems on VCTK-3mix with absolute F1-score improvements of 38.53\% and 4.14\%, respectively. We also achieve 39.96\% and 1.47\% absolute F1-score improvements on VCTK-4mix, respectively. The results demonstrate that the SSL task with more sources is more challenging, and validate that our ISSL system still maintains superiority on both VCTK-3mix and VCTK-4mix.


\begin{table}[t]
\centering
\small
\caption{Precision (\%), Recall (\%) and F1 (\%) in a comparative study of DOA estimation for unknown number of sources.}
\vspace{-5px}
    \setlength{\tabcolsep}{1.2mm}{
        \begin{tabular}{c|c|c|c|c}
        \toprule
        \textbf{Dataset} & \textbf{Methods} & \textbf{Precision}& \textbf{Recall} & \textbf{F1} \\
        \midrule
        \multirow{3}{*}{VCTK-3mix} & Threshold (CNN)   & 49.40 & 40.92 & 44.76 \\
        & Threshold (Res)   & 77.01 & 81.42 & 79.15 \\
        & Our ISSL &\textbf{84.41} &\textbf{82.19} &\textbf{83.29} \\
        \midrule
        \multirow{3}{*}{VCTK-4mix} & Threshold (CNN)  &45.40 &34.12 &38.96 \\
        & Threshold (Res) &77.87 &77.04 &77.45 \\
        & Our ISSL &\textbf{79.83} &\textbf{78.03} &\textbf{78.92} \\
		\bottomrule
        \end{tabular}}
\label{tbl:small_test_set}
\end{table}

\begin{table}[t]
\centering
\caption{Detection accuracy (\%) of the number of sound sources on VCTK-3mix and VCTK-4mix.}
    \setlength{\tabcolsep}{1.2mm}{
        \begin{tabular}{c|c|c}
        \toprule
        \textbf{Methods} &\textbf{VCTK-3mix}& \textbf{VCTK-4mix} \\
        \midrule
        Threshold (CNN) \cite{he2018deep} & 24.74  & 24.81  \\
        Threshold (Res) \cite{he2019adaptation} & 62.01  & 58.50  \\
        Our ISSL & \textbf{81.41}  & \textbf{75.45}  \\
        \bottomrule
        \end{tabular}}
\label{tbl:source_number_detection}
\end{table}



We further report the source number detection performance on VCTK-3mix and VCTK-4mix, as shown in Table \ref{tbl:source_number_detection}. It is observed that our ISSL system with iterative strategy achieves the best performance for detection of the number of sound sources. Specifically, our system leads up to 31.3\% and 29.0\% relative improvements than ``Threshold (Res)'' baseline system on both two databases. This observation further demonstrates that the performance improvements of DOA estimation mainly come from better detection on the number of sound sources.




\subsection{A universal study of ISSL system}
To prove the motivation that our ISSL system can deal with more sources than that seen during the training stage, we study a universal experiment summarized in Table \ref{tbl:universal_study}. As we can see in Table \ref{tbl:universal_study}, our ISSL system still works well when we directly use the model trained on VCTK-3mix to infer the results on VCTK-4mix. This can be explained that our system uses a data-driven way to produce the terminal criterion automatically, which allows the system to decide whether to continue iteration process and output the DOA result. These observed results verify the universal characteristics of our proposed ISSL system. In addition, we find that the performance of ISSL system is just slightly degraded. This phenomenon reflects our system's robustness, and also provides a solution to the problem of reducing training costs in SSL task.


\subsection{A stability analysis of ISSL system}
\begin{figure}[t] 
\centering 
\includegraphics[width=1.1\linewidth]{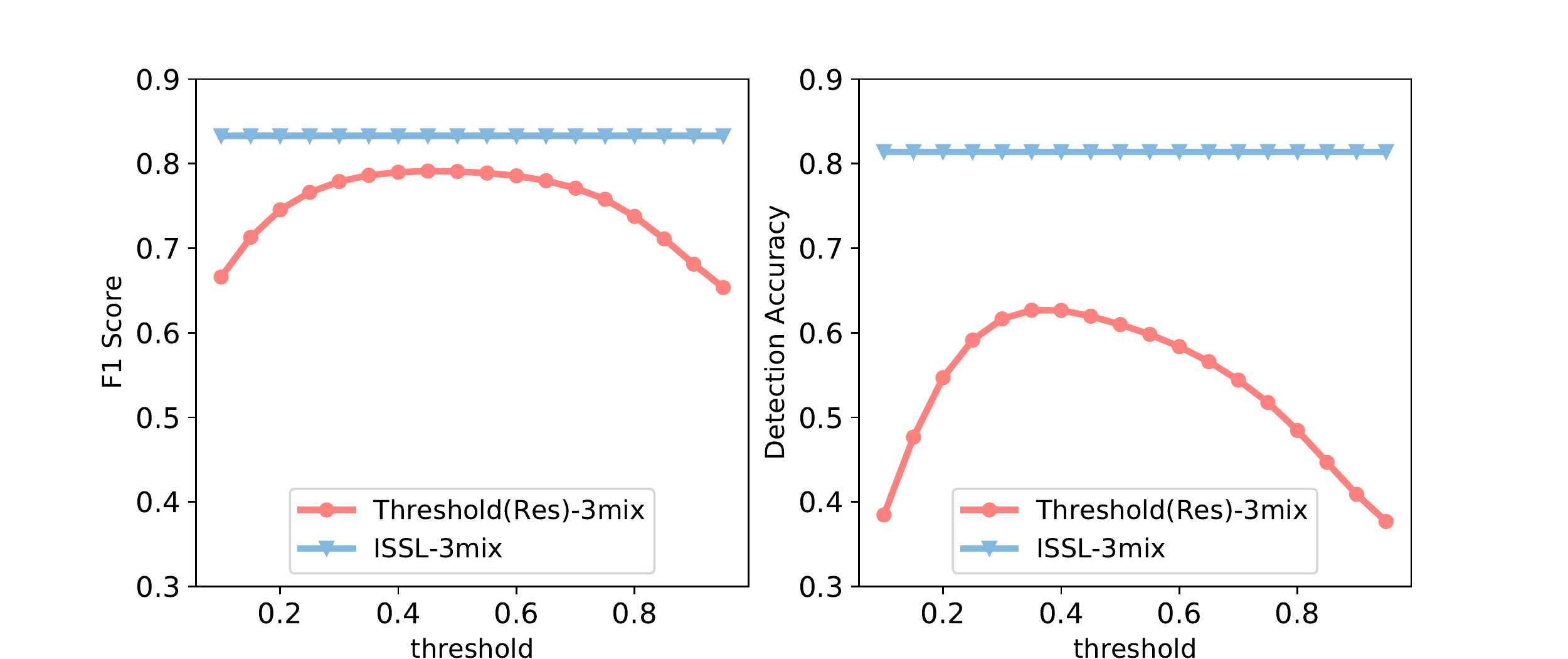} 
\caption{A stability analysis of ISSL and Threshold (Res) with different threshold values on VCTK-3mix. Left part is the results of DOA estimation, and right is source number detection results.
}
\label{Fig.Stability} 
\end{figure}

We further verify the robustness and stability of our ISSL system compared with the threshold-based baseline on different threshold values. The results are shown in Figure \ref{Fig.Stability}. We can conclude that the performance of DOA estimation and source number detection vary greatly with the change of the predefined threshold value, which is unsatisfactory for practical speech applications. Benefiting from the advantage of not requiring a threshold, our ISSL is quite stable.

\begin{table}[t]
\centering
\caption{A universal study on VCTK-4mix. Precision (\%), Recall (\%) and F1 (\%) are the metrics of DOA estimation. Detection accuracy (\%) is the metric of source number detection.}
    \setlength{\tabcolsep}{1.2mm}{
        \begin{tabular}{c|c|c|c|c|c|c}
        \toprule
        \multirow{2}{*}{\textbf{Model}} &\textbf{Train} & \textbf{Infer} &\multirow{2}{*}{\textbf{Precision}}& \multirow{2}{*}{\textbf{Recall}} & \multirow{2}{*}{\textbf{F1}} &\textbf{Detection}\\
        &\textbf{Data} & \textbf{Data} & & & &\textbf{Accuracy}\\
        \midrule
        \multirow{2}{*}{ISSL} &4mix & 4mix &79.83  &78.03  &78.92  &75.45 \\
        &3mix & 4mix &81.35  & 75.84  & 78.50 & 74.47  \\
        \bottomrule
        \end{tabular}}
\label{tbl:universal_study}
\vspace{-5pt}
\end{table}

\section{Conclusions}

We propose a novel iterative sound source localization approach to estimate DOA and detect the number of sound sources. Our novelty lies in designing an iterative strategy to replace the original threshold operation. This strategy allows our approach to handle an arbitrary number of sources, even more than the number of sources in the training step. Experimental results show that our proposed ISSL achieves better DOA estimation and source number detection performance for unknown number of sources  compared to the unstable threshold-based algorithms.


\bibliographystyle{IEEEtran}

\bibliography{mybib}


\end{document}